# Optical microcavity with semiconducting single-wall carbon nanotubes

Etienne Gaufrès,[1] Nicolas Izard,[1,2] Xavier Le Roux,[1] Saïd Kazaoui,[2] Delphine Marris-Morini,[1] Eric Cassan,[1] and Laurent Vivien[1,*]

[1]*Institut d'Electronique Fondamentale, CNRS-UMR 8622, Université Paris-Sud 11, 91405 Orsay, France*
[2]*National Institute of Advanced Industrial Science and Technology (AIST), 1-1-1 Higashi, Tsukuba, Ibaraki 305-8565, Japan*
[*]*laurent.vivien@u-psud.fr*

**Abstract:** We report studies of optical Fabry-Perot microcavities based on semiconducting single-wall carbon nanotubes with a quality factor of 160. We experimentally demonstrate a huge photoluminescence signal enhancement by a factor of 30 in comparison with the identical film and by a factor of 180 if compared with a thin film containing non-purified (8,7) nanotubes. Futhermore, the spectral full-width at half-maximum of the photo-induced emission is reduced down to 8 nm with very good directivity at a wavelength of about 1.3 µm. Such results prove the great potential of carbon nanotubes for photonic applications.

Semiconducting single wall carbon nanotubes (s-SWNT) have recently attracted a lot of interest due to their tunable direct band gap, allowing radiative electron-hole recombinaison, making them first-rate candidate for new optoelectronic and photonic applications[1-3]. Indeed, s-SWNT are extremely sensible to their environment and external fields. Their optical band gap could be tune, turning them from tunable light emitter to broad range light detector, allowing the emergence of a new all-carbon nanotube based photonics. Up to now, research was focused on the photoluminescence properties of individual or well isolated s-SWNTs, either by wrapping them with surfactant[4,5] or suspending them over trenches[6], with numerous impressive results achieved, such as single tube emitters[3], and recently isolated s-SWNTs integration into planar microcavity were also reported[7,8]. However, carbon nanotube doped thin films and their subsequent integration into optical devices was seldomly studied, with few results. While s-SWNTs display very strong photoluminescence properties, metallic carbon nanotubes (m-SWNTs), catalytic impurities and bundled s-SWNTs remaining in samples quenche photoluminescence by non radiative desexcitation paths, from m-SWNT interactions or s-SWNT exciton energy transfert[9]. The extraction of s-SWNT is a real challenge, and several ways have been explored in the last few years, such as functionalisation, DNA, polymer wrapping or gradient centrifugation[10-13]. We recently reported the selective extraction of a few s-SWNT chirality, without remaining trace of m-SWNT[14]. Fluorescence signal arising from thin polymer layers enriched in s-SWNT is highly enhanced compared to the usual mix of s- and m-SWNT[15]. Thanks to this breakthrough, we have now been able to make the following step towards the development of nanotubes based photonic devices, integrating this optically active medium into a photonic structure.

We report on the optical response of purely semiconducting SWNTs integrated into a Fabry-Perot planar cavity[16] constituted by two Bragg mirrors. Photoluminescence intensities of s-SWNT doped polymer thin films with and without its insertion into an optical Fabry-Perot cavity are compared. The (8,7) emission directionality across the cavity is studied and compared with literature results.

s-SWNT thin films were fabricated as previously described[14,15]. First, as-prepared HiPCO SWNT powder from Carbon Nanotechnologies Inc., poly-9,9-di-n-octyl-fluorenyl-2,7-diyl (PFO) from Sigma Aldrich and toluene were mixed in 1:4:6 ratio and homogenized by sonication for 1 h using a water-bath sonicator and 15 min using a tip sonicator. After 60 min centrifugation at 150000g, the upper 80 % of the supernatant solution was collected. This semiconducting SWNT purification step allows the enhancement of the photoluminescence signal by a factor of 6 for roughly the same s-SWNT species concentration. Indeed, as most of nanoparticles, m-SWNTs and nano-bundles are filtered, the background absorption is dramatically lowered and this opens interesting opportunities for photonic and optoelectronic applications[15]. The purified layer is then inserted between two Bragg mirrors to achieve Fabry-Perot cavity (Fig. 1). Bragg mirror consists of a multilayer-stack of alternate high and low refractive index films, all one quarter wavelength thick. Ten periods composed by successive alternation of silica ($SiO_2$, refractive index of 1.45) and silicon nitride ($Si_3N_4$, refractive index of 1.93) make up each Bragg mirrors. The $SiO_2$ and $Si_3N_4$ layer thicknesses are 215 nm and 160 nm respectively. Silica and silicon nitride layers are deposited on optical glass substrate by pressure enhanced chemical vapour deposition (PECVD) technique at 300 °C. The transmission spectrum of the Bragg mirror is reported in Fig. 2. Key features are the high reflectivity plateau centered at 1300 nm, designed to match the s-SWNT emission range, and the transparent range around 700-800 nm which correspond to the SWNT $E_{22}$ absorption bands, allowing a good optical pumping of the cavity. The plateau at 1300 nm has a reflectivity of more than 95 % in a 150 nm wavelength width.

The purified SWNT solution was then spin-casted on top of a dielectric Bragg mirror. The layer thickness was determined by ellipsometric spectroscopy, and was found to be around 200 nm. This result was later confirmed by the direct observation of a s-SWNT coated Bragg mirror slice by SEM (cf Fig. 1). The s-SWNT layer thickness is fundamental to adjust the cavity length, which control the resonance of the Fabry-Perot cavity. Ideally, the s-SWNT layer thickness should be adapted to precisely match the s-SWNT maximum emission peak[17], but this was proved to be technically challenging with our current setup. Lastly, the Bragg mirror was annealed at 150 °C for 15 min to obtain an homogeneous s-SWNT layer surface.

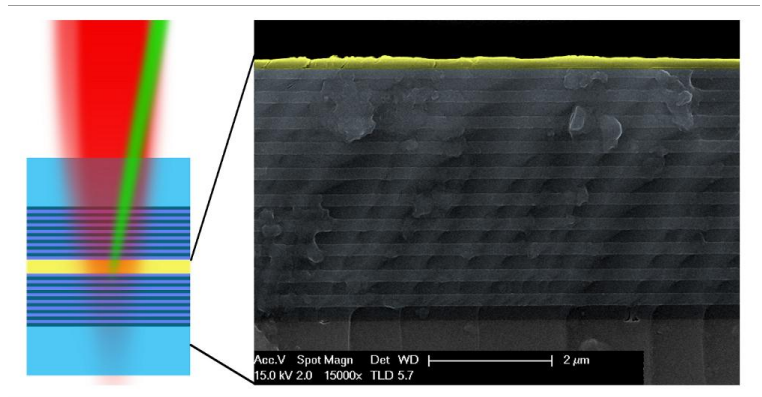

Fig. 1. Left: Schematic diagram of the Fabry-Perot microcavity constituted by two Bragg mirrors and the s-SWNT layer. Green(light) and red(dark) beams represent the pump and the emission beams, respectively. Right: SEM picture of the bottom Bragg mirror edge, displaying the alternate multilayer-stack of $SiO_2/Si_3N_4$ and the top s-SWNT layer.

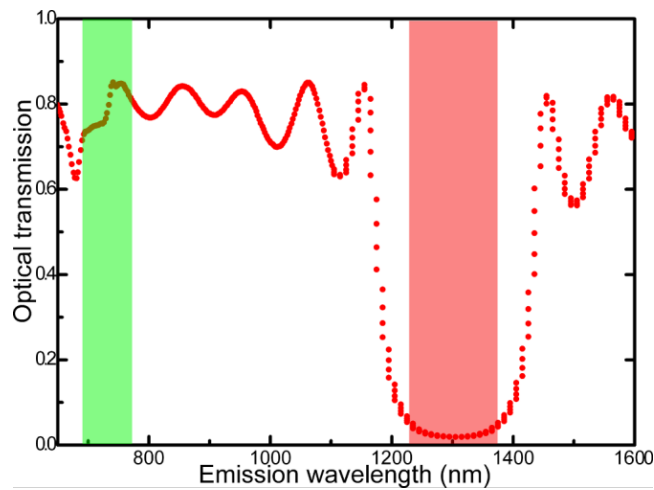

Fig. 2. Bragg mirror transmission spectrum. Green(light): Transparent region allowing pumping in the s-SWNT $E_{22}$ absorption bands. Red(dark): High reflectivity plateau matching s-SWNTs emission range.

The Fabry-Perot cavity is then obtained by bringing an upside-down Bragg mirror in contact to s-SWNT layer (cf. Fig. 1). The active thin layer photoluminescence is optically pumped using a Titane Sapphire (Ti:Sa) laser pumped by a cw Ar laser. Ti:Sa delivers continuous light at a wavelength range from 700 nm to 840 nm. The photoemission is collected and recorded at room temperature using a monochromator (JobinYvon 550) coupled with a cooled InGaAs detector with a 1.5 nm step. The numerical aperture of the collecting setup is 0.26, corresponding to a 15° collecting angle. First, photoluminescence map of the layer without cavity was recorded as a function of excitation and emission wavelengths. Fig. 3(a) shows the free emission peak of (8,7) SWNT around 1300 nm. This peak is very broad, with a full width half maximum (FWHM) of 40 nm. The same layer was then deposited in the Fabry-Perot resonator previously described and the photoluminescence measurement was performed in the same conditions. The optical response of the layer inside the cavity is presented in Fig. 3(b). We found only one Fabry-Perot resonance inside the reflectivity plateau. This is expected for a s-SWNT layer thickness thiner than twice the Bragg plateau broadness (i.e. < 300 nm). We could notice in the Fabry-Perot resonance an apparent second peak at $\lambda_{Ex} = 750$ nm, above the maximum emission peak of $\lambda_{Ex} = 740$ nm. This is caused by the fluctuation of the Bragg mirror transmission in this range, resulting in an incident pumping energy higher at 750 nm than at 740 nm. After correction of the Bragg mirror transmission in the pumping range, we obtain a single peak, centered at $\lambda_{Ex} = 740$ nm and $\lambda_{Em} = 1330$ nm (Fig. 3(c)). The observed Fabry-Perot resonance is strikingly thinner than the unconfined (8,7) SWNT emission peak, with a FWHM of 8 nm. The experimental cavity quality factor ($Q = \lambda/\Delta\lambda$) has been estimated around 160. This is so far the thinest resonance peak and the highest Q-factor reported using SWNT-based film resonators[7,8].

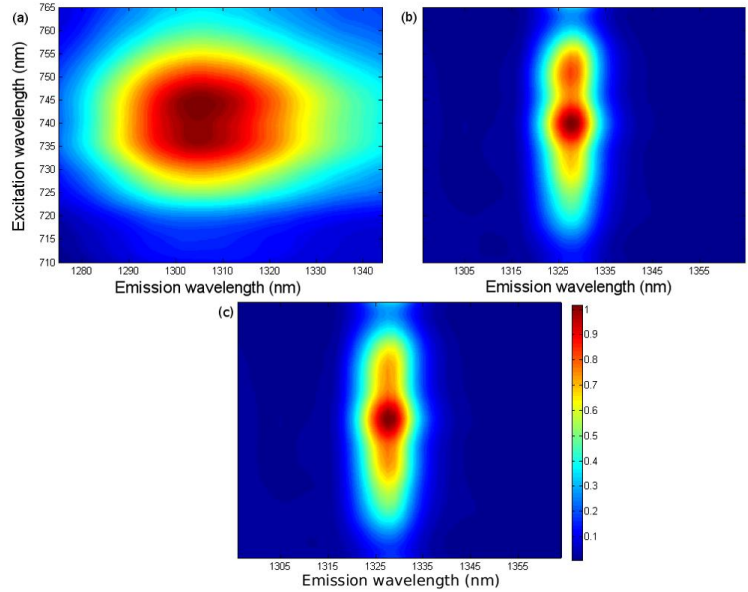

Fig. 3. Photoluminescence map of the s-SWNT thin layer : (a) without insertion into an optical cavity, (b) with insertion, and (c) after normalization by the Bragg mirror transmission. For clarity, each map intensity was normalized to 1.

The fluorescence intensity of s-SWNT thin layer after insertion into the Fabry-Perot cavity is presented in Fig. 4, for a fixed pumping wavelength of 740 nm. The fluorescence signal of the resonance is more than 30 times higher than the signal of the reference thin layer, demonstrating an interesting photoluminescence signal improvement using a single Fabry-Perot cavity. The signal was even enhanced by a factor 180 if we compare it with a standard non semiconducting enriched SWNT layer[15]. The cavity resonance peak position depends on the cavity length[17], and therefore could be adapted to match with the maximum emission peak of (8,7) SWNT and achieve still better signal enhancement. With our current setup, the cavity length is determined by the thickness of the spin-casted s-SWNT layer, but an integrated structure based on SWNT is currently being considered to adress this issue. However, it is remarkable that such a high signal amplification could be attained in this non optimized configuration. This proves the potential of such a structure to improve luminescence signal of SWNT using Fabry-Perot cavity, and open interesting perpective towards integrated devices and photonics application such as near-IR emitters.

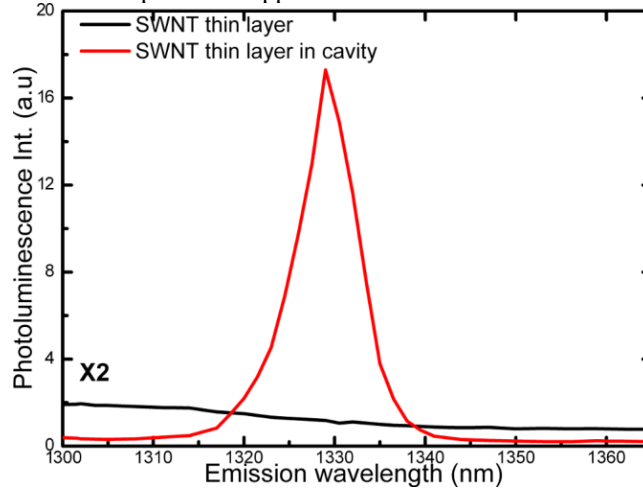

Fig. 4. s-SWNT thin layer fluorescence intensity as a function of cavity insertion under excitation at 740 nm. For clarity, the thin layer intensity was multiplied by a factor 2.

The directivity of the luminescence at the cavity output was also studied. Normalized photoluminescence intensity at 1330 nm of the s-SWNT layer inserted into the cavity is plotted in Fig. 5 as a function of the orientation angle between the cavity and the collector. As a reference, the angular dependence of the same layer in the half-cavity (upper mirror removed) is also reported. The cavity emission is indeed very selective in orientation, and less than 40 % of the original emission intensity is emitted outside a 6° cone, demonstrating a good confinement of the light inside the cavity.

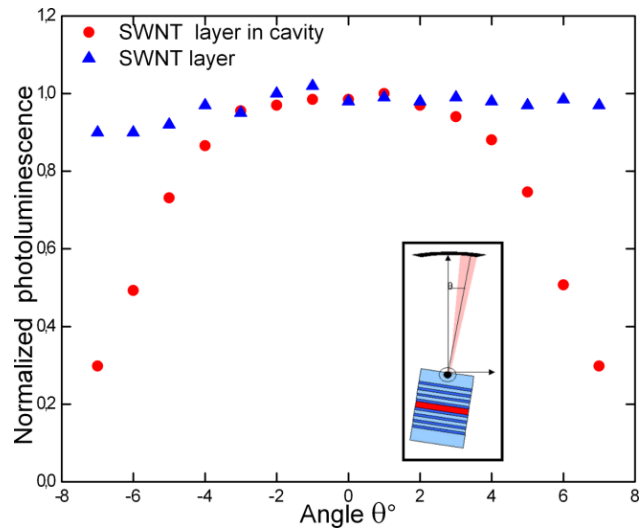

Fig. 5. Influence of the Fabry-Perot cavity orientation on the photoluminescence signal.

In conclusion, we demonstrated for the design of an optical Fabry-Perot cavity including a polymer layer highly doped in s-SWNT. The photoluminescence of s-SWNTs inside the cavity was studied, and we demonstrated a signal enhancement of more than 180 in comparaison with a non semiconducting enriched SWNT layer. The cavity quality factor of 160, and the emission directivity inside a 6° cone make it the actual state-of-the-art for a nanotube based cavity. Possible improvements using integrated structure on silicium, such as the selection of the nanotube emission wavelength, coupled with the strong emission enhancement using a Fabry-Perot cavity is the first step leading to all carbon nanotube-based photonics.

Authors acknowledge F.H. Julien and M. Tchernicheva for their help with photoluminescence experiments. N. Izard thanks the Japan Society for the Promotion of Science and CNRS for financial support.